\documentclass[%
 preprint,
superscriptaddress,
 amsmath,amssymb,
 aps,
]{revtex4-2}

\usepackage{graphicx}
\usepackage{dcolumn}
\usepackage{bm}
\usepackage[version=3]{mhchem}
\usepackage{gensymb}
\usepackage{setspace}

\begin{document}

\title{Two-gap to Single-gap Superconducting Transition on a \\ Honeycomb Lattice in \ce{Ca_{1-x}Sr_{x}AlSi}}

\author{Dorota~I.~Walicka}
\affiliation{Department of Chemistry, University of Zurich, CH-8057 Zurich, Switzerland}

\author{Zurab~Guguchia}
\affiliation{Laboratory for Muon Spin Spectroscopy, Paul Scherrer Institute, CH-5232 Villigen, Switzerland}

\author{Jorge~Lago}
\affiliation{Department of Chemistry, University of Zurich, CH-8057 Zurich, Switzerland}
\affiliation{Departamento de Química Inorgánica, Universidad del País Vasco, UVP/EHU, E-48080, Bilbao, Spain}

\author{Olivier~Blacque}
\affiliation{Department of Chemistry, University of Zurich, CH-8057 Zurich, Switzerland}

\author{KeYuan~Ma}
\affiliation{Department of Chemistry, University of Zurich, CH-8057 Zurich, Switzerland}

\author{Rustem~Khasanov}
\affiliation{Laboratory for Muon Spin Spectroscopy, Paul Scherrer Institute, CH-5232 Villigen, Switzerland}

\author{Fabian~O.~von~Rohr}
\email{To whom correspondence should be addressed.}
\affiliation{Department of Chemistry, University of Zurich, CH-8057 Zurich, Switzerland}
\affiliation{Department of Physics, University of Zurich, CH-8057 Zurich, Switzerland}


\date{\today}

\begin{abstract}

It is a well-established fact that the physical properties of compounds follow their crystal symmetries. This has especially pronounced implications on emergent collective quantum states in materials. Specifically, the effect of crystal symmetries on the properties of superconductors is widely appreciated, although the clarification of this relationship is a core effort of on-going research. Emergent phenomena on honeycomb lattices are of special interest, as they can give rise to spectacular phenomenology, as manifested by the recent discovery of correlated states in magic-angle graphene, or by the high-temperature superconductivity in \ce{MgB2}. Here, we report on the structural and microscopic superconducting properties of a class of ternary superconductors with Al/Si honeycomb layers, i.e. \ce{Ca_{1-x}Sr_{x}AlSi}. We show that this solid solution is a remarkable model system with a highly tunable two-gap to single-gap superconducting system on a honeycomb lattice, where the superconductivity is enhanced by a subtle structural instability, i.e. the buckling of the Al/Si layers.  \\

\end{abstract}

\raggedbottom                                
\maketitle


\newpage
\section{Introduction}\label{sec:introduction}

The increase of the superconducting transition temperature at a structural phase boundary is a widely observed behavior in various superconductors \cite{Slusky2001,Hirai2012,Lang1992,Torikachvili2008}. The relationship between structural instabilities and superconductivity has been a recurring theme in superconductivity research. These structural phase transitions are associated with a phonon softening, which influences the phonon-mediated Cooper pairing and consequentially may lead to a enhancement of the superconductivity \cite{Allen1972,Wu2018}. 
In particular, many of the known superconductors are chemically and electronically located close to structural phase instabilities, e.g. \ce{Nb3Sn}-related, bismuth-oxide, or graphite intercalated superconductors \cite{Testardi1975,Hinks1988,Gauzzi2007}.

The recent discovery of superconductivity in magic-angle bi-layer graphene has reignited the interest in superconductors with honeycomb lattices \cite{Cao2018}. The most prominent bulk superconductor with (boron) honeycomb layers sandwiched between Mg(II) layers, is \ce{MgB2} with a critical temperature of $T_{\rm c}$ = 39 K \cite{Nagamatsu2001}. This remains the highest critical temperature at ambient pressure to-date, among materials, in which Cooper pairing is believed to be mediated by phonons, and described by the BCS (Bardeen–Cooper–Schrieffer) theory \cite{Bardeen1957}. The extraordinary superconducting properties of \ce{MgB2} are widely believed to be caused by two types of electrons at the Fermi level with widely differing behaviors leading to the opening of two superconducting gaps \cite{Mazin2003, Canfield2001}. 

Other honeycomb bulk superconductors are intercalated variants of graphite. Prominent examples of this group are \ce{CaC6} with a critical temperature of $T_{\rm c}$ = 11~K, which increases up to 15.1~K at a pressure of $p =$~8~GPa \cite{Gauzzi2007} and \ce{YbC6} with a critical temperature of $T_{\rm c}$ =~6.5~K \cite{Weller2005}. In this case the superconductivity is likely caused by intercalant phonons \cite{Mazin2005GIC,Calandra2005GIC} or by an enhanced electron–phonon coupling through acoustic plasmons \cite{Csanyi2005GIC}.
 
Besides these binary superconductors, there are only few ternary honeycomb-based superconductors known. Most recently, SrPtAs with a critical temperature of $T_{\rm c} \approx$ 2.4 K has been of great interest, since it was predicted to be the first chiral, and therefore first intrinsic topological superconductor with a $d + id$ gap on a layered hexagonal lattice  \cite{Nishikubo2011,Biswas2013,Fischer2014}. Among the ternary honeycomb-based superconductors has the MAlSi (M= Ca, Sr, Ba) family been most prominently discussed, with their critical temperature of up to $T_{\rm c}$ = 7.8 K \cite{Imai2002,Imai2003,Lorenz2002,Lorenz2003,Nakagawa2006,Yamanaka2007,Meng2003,Ma2008,Evans2009}. SrAlSi and BaAlSi crystallize in the AlB\textsubscript{2}-type structure with Al/Si honeycomb layers, which are intercalated by the earth-alkaline metal \cite{Lorenz2003,Nakagawa2006,Evans2009}. The structure of CaAlSi is related to the \ce{AlB2} structure, but more complex. It is believed that this material can crystallize in few polymorphic structures: denoted as 1H, 5H, and 6H structures, here H stands for hexagonal and the digit represents the amount of honeycomb Al/Si layers in the unit cell \footnote[1]{Whenever the chemical formula CaAlSi is used throughout this manuscript, then we always talk about the 6H polymorph, since this is the thermodynamic stable stoichiometric form, if not otherwise explicitly noted.}. The critical temperatures vary between the different CaAlSi polymorphs between $T_{\rm c}$ = 5.7 to 7.8 K, however, all of them were reported to consist partly of planar honeycomb Al/Si layers \cite{Kuroiwa2006,Sagayama2006}. CaAlSi and SrAlSi superconductors are widely believed to be conventional phonon-mediated superconductors. There are, however, also reports that indicate uncommon superconducting properties, such as e.g. unconventional pressure effects \cite{Lorenz2003, Boeri2008} or field dependent muon spin rotation ($\mu$SR) measurements on CaAlSi \cite{Kuroiwa2007}.

Especially noteworthy are a series of detailed specific heat measurements, which is indicated different mechanisms behind the superconductivity in CaAlSi and SrAlSi \cite{Lorenz2003}. In this work, \textit{B. Lorenz et al.} suggested that even though, these two compounds are isoelectronic and belong to the same structural family, their properties differ significantly: SrAlSi can be best described by a weak coupling BCS theory and has negative response on applied pressure, while CaAlSi was found to be best described by a strong coupling BCS theory and has a positive pressure coefficient in the low-pressure ($p <$ 1.2 GPa) regime. Moreover, both of them were predicted to be one gap superconductors in contrast to MgB\textsubscript{2} \cite{Lorenz2004,Tsuda2004}.

Here, we investigate the structural and the microscopic superconducting properties of the \ce{Ca_{1-x}Sr_{x}AlSi} solid solution. Specifically, we have realized the continuous solid solution, which for all members, other than $x =$~0, crystallize in the \ce{AlB2}-type structure. We present an improved structural model for the parent compound CaAlSi by means of single-crystal X-ray diffraction. We find that in opposition to the \ce{AlB2} structure, all honeycomb-type Al/Si layers of CaAlSi are slightly buckled. Our results strongly indicate that the structural instability in CaAlSi enhances the superconductivity across the solid solution. This is further supported by an investigation of the London magnetic penetration depths ${\lambda}^{-2}$, where we find strong evidence for the transition of a two-gap $s+s$-wave superconducting gap to a single-gap $s$-wave model, which we find to coincide with the disappearance of the structural distortion across the solid solution.

\section{Methods}\label{sec:exp}
\textbf{Synthesis:} All compounds were synthesized from the pure elements (Ca 99.99 \%  Aldrich; Sr 99.99 \%  Aldrich; Al 99.9995 \% Acros Organic; Si 99.95 \%  Aldrich). Ca and Sr were handled inside an argon glovebox to prevent the oxidation of the reactants. Stoichiometric amounts of the metals were arc-melted in an argon atmosphere on a water-cooled copper plate (anode) with the use of a tungsten tip (cathode). A Zr sponge was co-heated in the system to further purify the reaction atmosphere from oxygen or water. In case of SrAlSi, a 5\% excess of strontium was used for the highest purity product. The metals were melted together and remelted three times, each time flipping the sample over, in order to provide a maximized homogeneity of the prepared sample. Single crystals of CaAlSi, \ce{Ca_{0.8}Sr_{0.2}AlSi}, \ce{Ca_{0.6}Sr_{0.4}AlSi}, \ce{Ca_{0.5}Sr_{0.5}AlSi} and SrAlSi were prepared by slowly decreasing the applied voltage over a few seconds.

\textbf{Diffraction:} Powder X-ray diffraction (PXRD) patterns were obtained on a STOE STADIP diffractometer equipped with a Ge-monochromator using Cu-K\textsubscript{$\alpha$1} radiation ($\lambda$ = 1.54051 \r{A}). All patterns were measured in the 5-90$^{\circ}$ 2$\Theta$ range with a scan step of 0.015$^{\circ}$. The cell parameters were obtained by performing LeBail fits using the FULLPROF program package \cite{Rodriguez1993}. Single crystal X-ray diffraction (SXRD) data were collected on a Rigaku OD Synergy (Pilatus 200K Hybrid Pixel Array detector) diffractometer for single crystal from nominal compositions SrAlSi, \ce{Ca_{0.8}Sr_{0.2}AlSi}, \ce{Ca_{0.6}Sr_{0.4}AlSi}, \ce{Ca_{0.5}Sr_{0.5}AlSi}, and on a Rigaku OD SuperNova (Atlas CCD detector) diffractometer for CaAlSi, both equipped with an Oxford liquid-nitrogen cryostream cooler. A single wavelength X-ray source from a micro-focus sealed X-ray tube was used for the analyses (Cu-K\textsubscript{$\alpha$} radiation, $\lambda$ = 1.54184 \r{A}). Suitable single crystals were manipulated into polybutene oil, mounted on a flexible loop fixed on a goniometer head and transferred to the diffractometer. Data collections, data reductions, and analytical absorption corrections \cite{Clark1995} were performed with the program suite \textit{CrysAlisPro}. Using \textit{Olex2} \cite{Dolomanov2009}, the structures were solved with the SHELXT small molecule structure solution program \cite{Sheldrick2015} and refined with the SHELXL program package \cite{Sheldrick2015a} by full-matrix least-squares minimization on F\textsuperscript{2}.

\textbf{Physical Properties:} Magnetic measurements were carried out on a Quantum Design Magnetic Properties Measurement System (MPMS XL) equipped with a reciprocating sample option (RSO) and a 7 T magnet. The ${\mu}$SR experiments were carried out at the Swiss Muon Source (S${\mu}$S) Paul Scherrer Insitute, Villigen, Switzerland using the low background GPS (${\pi}$M3 beamline) instrument, the high field HAL-9500 ${\mu}$SR spectrometer (${\pi}$E3 beamline), equipped with BlueFors vacuum-loaded cryogen-free dilution refrigerator (DR) and high pressure GPD instrument (${\mu}$E1 beamline), equipped with \emph{Oxford~Instruments~Heliox} $^{3}$He cryostat. For the experiment on the HAL-9500 instrument, a pellet with a diameter of 10 mm was used. For pressure experiments, three pellets of 5.9 mm in diameter were used.
Pressures up to 1.9 GPa were generated in a double wall piston-cylinder type of cell made of MP35N material, especially designed to perform ${\mu}$SR experiments under pressure \cite{GuguchiaPressure,MaisuradzePC,GuguchiaNature}. As a pressure transmitting medium Daphne oil was used. The pressure was measured by tracking the superconducting transition of a very small indium plate by AC susceptibility. The filling factor of the pressure cell was maximized. The fraction of the muons stopping in the sample was approximately 40 ${\%}$. The ${\mu}$SR time spectra were analyzed using the free software package MUSRFIT \cite{Suter2012}.\\

\section{Results and discussion}\label{sec:results}

\subsection{Crystal Structure and Phase Purity }\label{sec:structure}

\begin{figure}
	\includegraphics[width=1\linewidth]{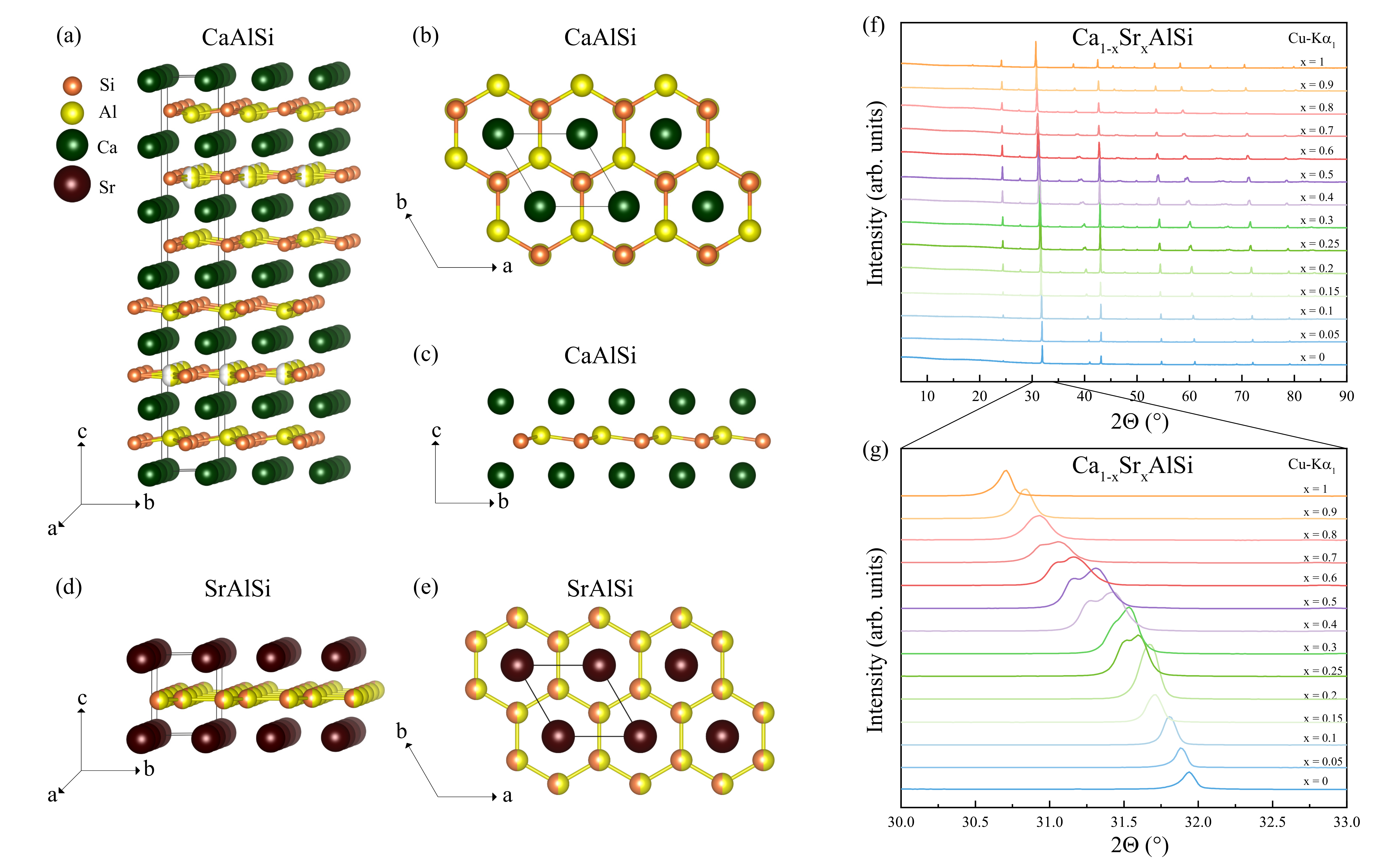}
	\caption{Structure and bonding of the Ca\textsubscript{1-x}Sr\textsubscript{x}AlSi solid solution.  Crystal structure of CaAlSi with space group \textit{P}6\textsubscript{3}\textit{/mmc} (a) along the $a$ direction with an emphasis on the 6-layered super-cell, (b) along the $c$ direction, showing the honeycomb Al/Si layers, and (c) a single layer of CaAlSi showing the buckling of the Al/Si layers. Crystal structure of SrAlSi with space group \textit{P}6\textit{/mmm} in the \ce{AlB2} structure type (d) along $a$ direction and (e) along $c$ direction. (f) Powder X-ray diffraction patterns of the Ca\textsubscript{1-x}Sr\textsubscript{x}AlSi solid solution with $x$ = 0, 0.05, 0.1, 0.15, 0.2, 0.25, 0.3, 0.4, 0.5, 0.6, 0.7, 0.8, 0.9, and 1 (g) Zoom-in of 2$\Theta$ = 30.0 to 33.0 region showing the pronounced shifting as a function of Ca/Sr content and the slight splitting for the samples $x =$ 0.3 to 0.7 of the (101) Bragg reflection.}
	\label{fig:CaAlSistructure}
\end{figure}

We have synthesized the Ca\textsubscript{1-x}Sr\textsubscript{x}AlSi solid solution for $x$ = 0, 0.05, 0.1, 0.15, 0.2, 0.25, 0.3, 0.4, 0.5, 0.6, 0.7, 0.8, 0.9, and 1. All resulting products were found to be highly crystalline. For the end member $x$ = 0 of the solid solution, CaAlSi, we were able to obtain high quality single crystals with sizes up to 0.02 x 0.04 x 0.06 mm$^3$ by this method. These allowed us to perform a detailed analysis of its crystal structure by means of SXRD. Earlier result from powder synchrotron and neutron diffraction experiments suggested that the unit cell of CaAlSi should to be 6-fold compared to the unit cell of the \ce{AlB2} structure \cite{Sagayama2006}. \textit{Sagayama et al.} had earlier developed a structural model, where they proposed that some of the honeycomb Al/Si layers in CaAlSi are buckled, while others remained planar. 

Our detailed analysis of the crystal structure shows now that: (i) the aluminum and silicon atoms are indeed not occupying random positions as it would be required for the \ce{AlB2} structure, and (ii) the unit cell is indeed 6-fold along the $c$-direction. (iii) Our data, however, unequivocally shows that all of the honeycomb Al/Si layers are in fact buckled and that none of them remain planar, which is in contrary to earlier structural models. The aluminum atoms are thereby randomly distributed above and below the plain in in the second and fifth layer of the unit cell. The distortion of the structure, which results in the buckling and different stacking of the Al/Si layers result in the space group \textit{P}6\textsubscript{3}\textit{/mmc}. It is worth noticing that the displacement within the Al/Si layers in CaAlSi is small. The silicon atoms remaining in the plain and the aluminium atoms are displaced by $\Delta r$ of 0.173(11) \AA. This structure resembles the well-known YPtAs-type structure with 4 hexagonal buckled layers, but the displacement of some of the aluminum atoms below the plane are causing CaAlSi to crystallize in its own, unique structure-type. The structure of CaAlSi with the unit cell parameters of a~=~4.1812(2)~{\AA} and c~=~26.3009(12)~{\AA} is shown in figure \ref{fig:CaAlSistructure}, the structure file has been deposited with the number CSD-2013753 in the CCDC/FIZ Karlsruhe databases. In figure \ref{fig:CaAlSistructure}(b) the view along the $c$ direction is presented, which reveals how the hexagonal honeycomb Al/Si layers are stacked between layers of calcium atoms. The summary of the SXRD refinement are shown in the table \ref{tab:crystal} and the details of the crystallographic positions are presented in table \ref{tab:Wyckoff}. The aluminium atoms occupy the \textit{4f} Wyckoff position with occupancy 1 and 0.5, respectively, for those atoms who display disorder. The calcium atoms occupy the \textit{2a} and \textit{4e} positions, while Si atoms are located on the \textit{4f} and \textit{2c} positions. 

We find that all other synthesized members of the Ca\textsubscript{1-x}Sr\textsubscript{x}AlSi solid solution crystallize in the \ce{AlB2}-type structure with the \textit{P}6\textit{/mmm} space group, with random site occupancies in the Al/Si and Ca/Sr layers, respectively. These structures are confirmed by high-quality single-crystal data for samples with $x$ = 0.2, 0.4, 0.5 and 1 (the details of the refinements are listed in the Supplemental Information), the structure files have been deposited with the numbers CSD-2013750, CSD-2013751, CSD-2013752, and CSD-2013754 in the CCDC/FIZ Karlsruhe databases. The crystal structure, as a representative of the whole solid solution, of SrAlSi is depicted in figure \ref{fig:CaAlSistructure}(d)\&(e) along the $a$ and the $c$ axis, respectively.
 
In figure \ref{fig:CaAlSistructure}(f) the PXRD pattern of all prepared members of the Ca\textsubscript{1-x}Sr\textsubscript{x}AlSi solid solution are shown. The analysis of the patterns reveals that all reflections shift systematically toward smaller 2$\Theta$ angles. As can be expected, this corresponds to an increase of the unit cell parameters with increasing strontium content. All calcium containing samples have small impurities of the calcium deficient phase \ce{CaAl2Si2} and the calcium-rich phase \ce{Ca3Al2Si2}, accordingly. The impurities increase for synthesis attempts with alkaline-earth excesses, or upon annealing of the samples at various temperatures in quartz or niobium tubes. 

A zoom-in of the PXRD patterns is show in \ref{fig:CaAlSistructure}(g) in the 30-33$^{\circ}$ 2$\Theta$ range for the (101) Bragg reflection. The pronounced shifting of the (101) reflection can be clearly observed, furthermore a slight splitting of the reflection for $x$ values between 0.25 to 0.7 becomes apparent. The slight splitting is observed for the reflections (101), (201), (112), and (211). This is likely due to a lack of complete randomness in the distribution of calcium and strontium atoms along certain directions. It is unlikely that it is due to the presence of two phases with similar stoichiometries, because in this case two superconducting transitions close to each other would be expected. However the superconducting transitions of this compositions are remarkably sharp (see discussion below). Reheating of the samples in the arc furnace removed the splitting but at the same time significantly increases the  intensity of impurity peaks. Annealing and or reheating also substantially lowers the critical superconducting temperatures (see, Supplemental Information). We have performed LeBail fits for all the compounds in order to obtain the respective cell parameters. The cell parameters $a$ and $c$ are found to change nearly linearly, following Vegard's law.
 

\begin{table}
\begin{singlespace}
    \centering
    \begin{tabular}{|c|c|}
        \hline
 		\multicolumn{2}{|c|}{Single crystal data for CaAlSi} \\
 		\hline
Composition & CaAlSi \\
 CCDC/FIZ   & CSD-2013753 \\
Formula weight & 285.45 \\
Temperature/K &	160(1)  \\
Crystal system & hexagonal \\
Space group & \textit{P}6\textsubscript{3}/\textit{mmc} \\
a/\r{A} & 4.1812(2) \\
b/\r{A} & 4.1812(2) \\
c/\r{A} & 26.3009(12) \\
$\alpha$/\degree & 90 \\
$\beta$/\degree & 90 \\
$\gamma$/\degree & 120 \\
Volume/\r{A}\textsuperscript{3} & 398.20(4) \\
Z &	2 \\
$\rho$\textsubscript{calc} g/cm\textsuperscript{3} & 2.381\\
$\mu$/mm\textsuperscript{-1} & 24.847\\
F(000) & 282.0 \\
Crystal size/mm\textsuperscript{3} &  0.02 x 0.04 x 0.06 \\
Radiation &	Cu-K$\alpha$ ($\lambda$ = 1.54184 \r{A}) \\
2$\Theta$ range for data collection/\degree & 6.72 to 136.40 \\
Index ranges & -5 $\leq$ h $\leq$ 4 \\
& -4 $\leq k$ $\leq$ 5 \\
& -31 $\leq$ l $\leq$ 31 \\
Reflections collected &	181 \\
Independent reflections & 181 [R\textsubscript{int} =0.0082 \\
& R\textsubscript{sigma} = 0.0036] \\
Data/restraints/parameters & 181/0/18 \\
Goodness-of-fit on F\textsuperscript{2} & 1.221 \\
Final R indexes [I $\geq$ 2 $\sigma$ (I)] & R\textsubscript{1} =  0.0153 \\
& wR\textsubscript{2} = 0.0517 \\
Final R indexes [all data] & R\textsubscript{1} = 0.0187 \\
& wR\textsubscript{2} = 0.0545 \\
Largest diff. peak/hole / e Å\textsuperscript{-3} & 0.23/-0.16 \\
\hline
    \end{tabular}
    \caption{Details of the SXRD measurements and structural refinement for CaAlSi.}
    \label{tab:crystal}
    \end{singlespace}
\end{table}
   

 \begin{table}
     \centering
     \begin{tabular}{|ccccccc|}
     \hline
 		\multicolumn{7}{|c|}{Wyckoff positions} \\
 		\hline
 		Atom & Symbol & x & y & z & U\textsubscript{ISO} & Occ. \\ \hline
 		Al1 & \textit{4f} & 2/3 & 1/3 & 0.40928(4) & 0.0156(3) & 1\\
 		Al2 & \textit{4f} & 2/3 & 1/3 & 0.2566(4) & 0.0251(17) & 0.5\\
 		Ca1 & \textit{2a} & 0 & 0 & 1/2 & 0.0132(3) & 1\\
 		Ca2 & \textit{4e} & 0 & 0 & 0.33227(2) & 0.0151 & 1\\
 		Si1 & \textit{4f} & 1/3 & 2/3 & 0.42205(3) & 0.0149(3) & 1\\
 		Si2 & \textit{2c} & 1/3 & 2/3 & 1/4 & 0.0113(3) & 1\\ 
 		
 	\hline
    \end{tabular}
    \caption{Atomic coordinates, isotropic displacement parameters, and occupancies of the atoms in CaAlSi obtained by SXRD.}
    \label{tab:Wyckoff}
 \end{table}{}{}
 

\subsection{Magnetic measurement}\label{sec:magnetic}
  
\begin{figure}
\centering
\includegraphics[width=0.6\linewidth]{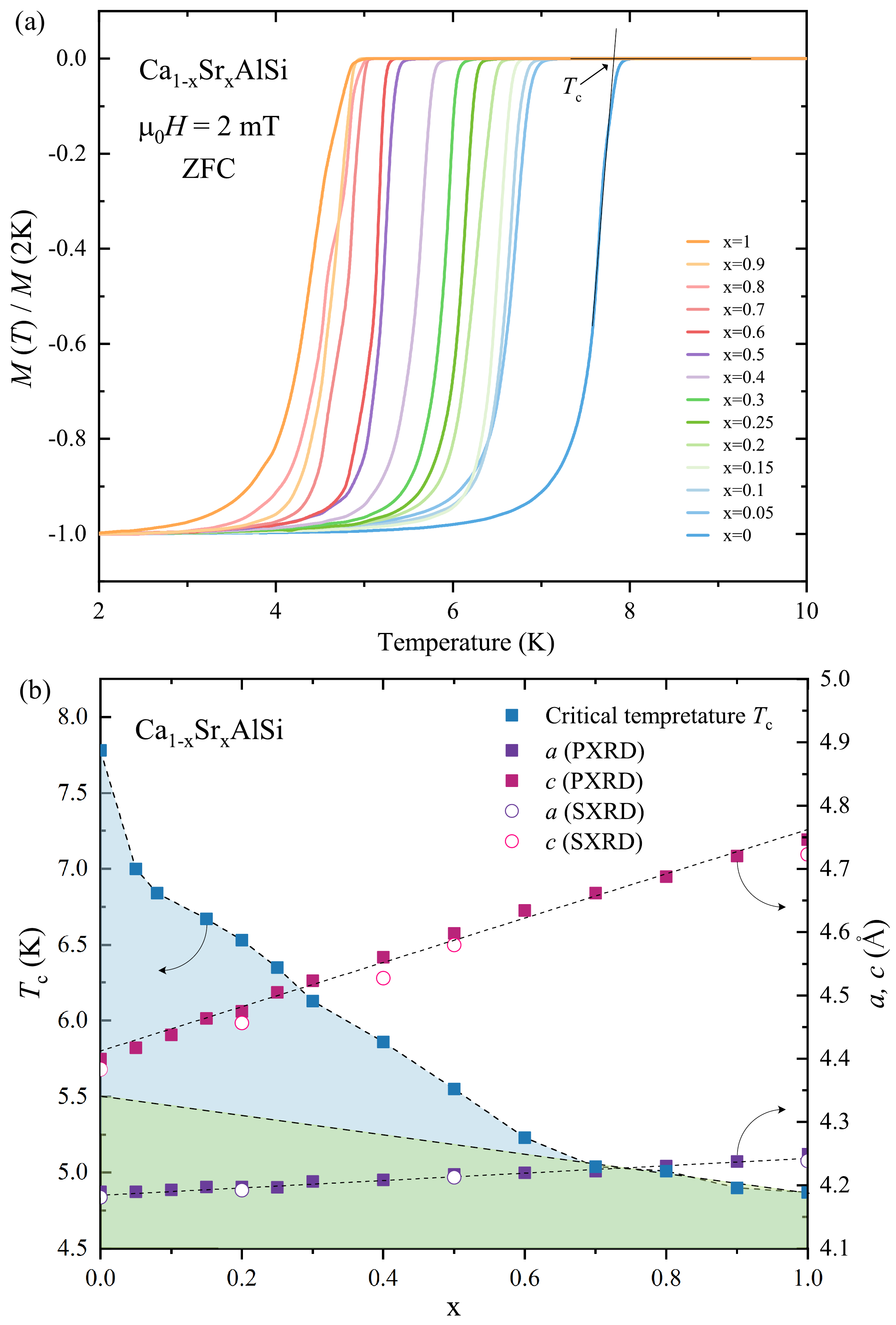}\vspace{-0.5cm}
\caption{Superconducting properties of the \ce{Ca_{1-x}Sr_xAlSi} solid-solution. (a) ZFC magnetization in an external magnetic field of $\mu_0 H = $ 2 mT for $x$ = 0, 0.05, 0.1, 0.15, 0.2, 0.25, 0.3, 0.4, 0.5, 0.6, 0.7, 0.8, 0.9, and 1, plotted between 1.75 and 10 K. Data are normalized by being plotted as $-M(T)/M({\rm 2K})$. (b) Summary of the crystallographic and superconducting parameters of the samples of the solid solution for $x$ = 0, 0.05, 0.1, 0.15, 0.2, 0.25, 0.3, 0.4, 0.5, 0.6, 0.7, 0.8, 0.9, and 1. The cell parameter $c$ for CaAlSi was divided by 6 for comparability. The unit cell values from the SXRD data are shown as circles, the ones from PXRD are shown as squares.}

\label{fig:squid}
\end{figure}

The zero-field-cooled (ZFC) magnetization of all prepared members of the solid-solution \ce{Ca_{1-x}Sr_xAlSi} with $x$ = 0, 0.05, 0.1, 0.15, 0.2, 0.25, 0.3, 0.4, 0.5, 0.6, 0.7, 0.8, 0.9, and 1 in an external field of $\mu_0 H$ = 2 mT in a temperature range between $T$ = 1.75 K and 10 K are depicted in figure \ref{fig:squid}. All magnetic shielding fractions were found to be larger than 100 \% for all the samples, due to demagnetization effects, confirming the bulk nature of the superconductivity in all samples. For better comparison, the magnetization data was normalized by plotting it as $-M(T)/M({\rm 2K})$. The transitions to the superconducting state are remarkably sharp for all the samples. The highest superconducting transition is observed for the parent compound of the solid solution $x$ = 0 (CaAlSi), with a critical temperature of $T_{\rm c} \approx$ 7.8 K. This value corresponds to the highest critical temperature reported for any polymorph of CaAlSi earlier (see, e.g., references \cite{Kuroiwa2006, Sagayama2006, Evans2009}). We find the critical temperature to monotonically decrease with increasing strontium content, reaching a value of $T_{\rm c} \approx$ 4.9 K for SrAlSi (x = 1). The measured transition temperature for the prepared samples by arc-melting of SrAlSi agrees well with previous reports (see, e.g., references \cite{Lorenz2003,Evans2009}). In figure \ref{fig:squid}, we have summarized the structural and superconducting parameters of the solid solution. It should be noted that the critical temperature in the solid solution does not follow the nearly linear change of the cell parameters, but that it has a more complex monotonic change as a function of strontium content. The critical temperature decreases more drastically close to the structurally distorted end member CaAlSi. This indicates that the higher $T_{\rm c}$ of CaAlSi and compositions close to it, are affected by an enhanced electron-phonon coupling due to a phonon-softening close to the  structural instability. A mechanism that can emerge in superconductors close to structural phase transitions.

\subsection{London magnetic penetration depth}

\begin{figure}
\centering
\includegraphics[width=0.5\linewidth]{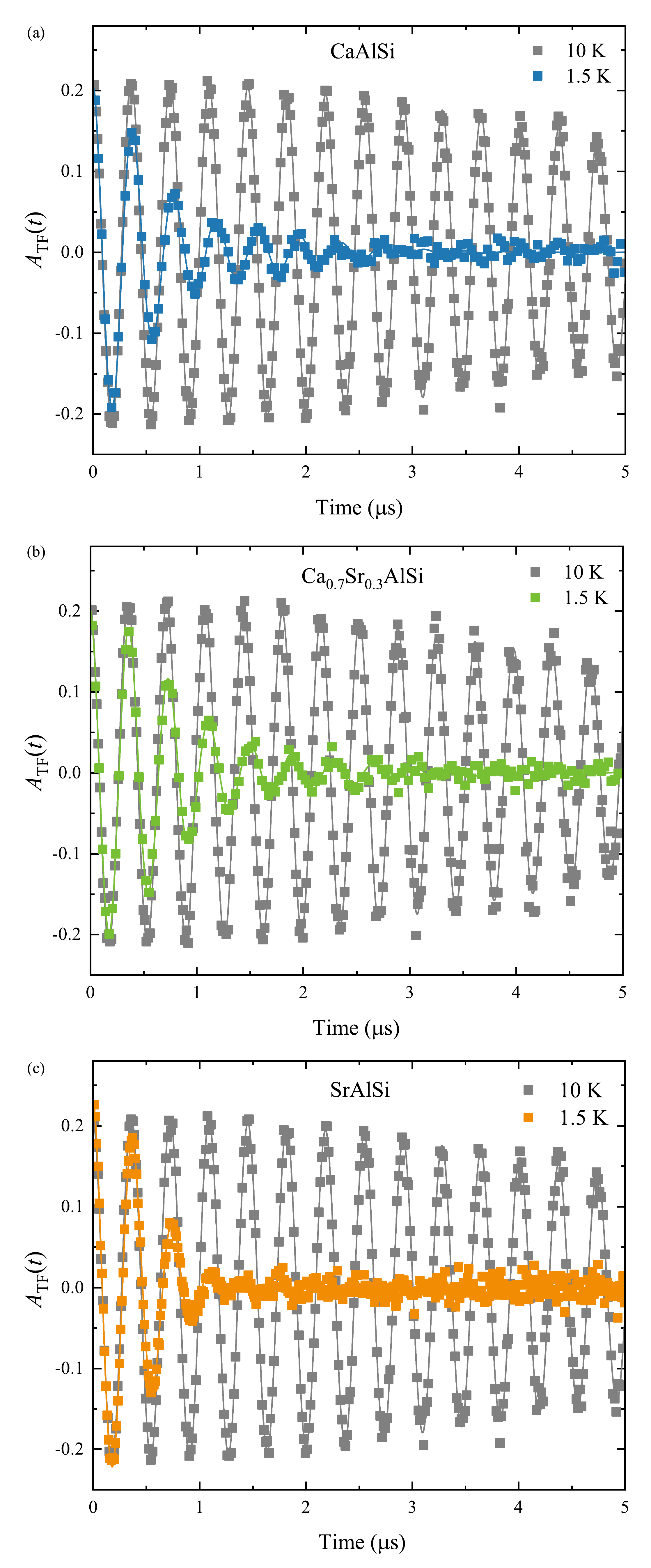}
\vspace{-0.5cm}
\caption{Transverse-field $\mu$SR time spectra of the \ce{Ca_{1-x}Sr_xAlSi} solid solution. (a) CaAlSi, (b) Ca$_{0.7}$Sr$_{0.3}$AlSi and (c) SrAlSi spectra in the normal-state at $T =$ 10 K (gray points) and in the superconducting state at $T =$ 1.5 K (blue, green, and orange points, respectively). The solid lines represent fits to the data points by means of equation (\ref{eq3}).}
\label{fig:muon3}
\end{figure}

We have used $\mu$SR measurements in order to investigate the London magnetic penetration depth in a temperature range between $T =$ 300 mK and 10 K. In ${\mu}$SR experiments nearly 100 \% spin-polarized muons ${\mu}^{+}$ are implanted into the investigated sample one at a time. The positively charged ${\mu}^{+}$ thermalize at interstitial crystallographic sites, where they act as magnetic microprobes. In a magnetic material the muon spin precesses in the local field $B_{\rm \mu}$ at the muon site with the Larmor frequency:

\begin{equation}
{\nu}_{\rm \mu} = \gamma_{\rm \mu}/(2 {\pi})B_{\rm \mu}.    
\end{equation}

For the muon the gyromagentic ratio is $\gamma_{\rm \mu}$/(2${\pi}$) = 135.5~MHz~T$^{-1}$. Using the $\mu$SR technique fundamental microscopic parameters of superconductors can be measured, namely the magnetic penetration depth $\lambda$ and the coherence length $\xi$ \cite{Sonier2000}.

If a type-II superconductor is cooled below critical temperature $T_{\rm c}$ in an applied magnetic field ranged between the lower ($H_{c1}$) and the upper ($H_{c2}$) critical fields, a vortex lattice is formed which in general is incommensurate with the crystal lattice with vortex cores separated by much larger distances than those of the unit cell. Because the implanted muons stop at given crystallographic sites, they will randomly probe the field distribution of the vortex lattice. These measurements are performed in a field applied perpendicular to the initial muon spin polarization (so called TF configuration) \cite{Sonier2000,Brandt1988,Blundell1999}.

In figure \ref{fig:muon3}, we present transverse-field (TF-$\mu$SR) time spectra measured above (10 K) and below (1.5 K) the superconducting transition temperature of (a) CaAlSi, (b) Ca$_{0.7}$Sr$_{0.3}$AlSi ($x$ = 0.3) and (c) SrAlSi, respectively. For all measured samples, oscillations at 10 K occur at a very small relaxation with time. This very small relaxation is due to random local magnetic fields within the samples. For the time spectra at 1.5 K the oscillations show a significant decrease due to the formation of the flux-line lattice (FLL), and thus nonuniform magnetic field distribution in the Shubnikov phase. The solid lines represent fits to the data points by using the following functional form \cite{Suter2012}:
\begin{equation}
\begin{aligned}
P_{s}(t)= \sum_{i=1}^{2} A_{s,i}\exp\Big[-\frac{\sigma_{i}^2t^2}{2}\Big]\cos(\gamma_{\mu}B_{int,s,i}t+\varphi). \\ 
\label{eq3}
\end{aligned}
\end{equation}
Two component expression was used for the sample due to the observed weak asymmetric field distribution. In equation \ref{eq3}, $A_{\rm s,i}$, $B_{\rm int,i}$ and ${\sigma}_{\rm i}$ are the asymmetry, the mean field and the relaxation rate of the i-th component, and ${\varphi}$ is the initial phase of the muon-spin ensemble. $\gamma/(2{\pi})\simeq 135.5$~MHz~T$^{-1}$ is the muon gyromagnetic ratio. In order to extract the second moment of the field distribution from the two component fitting we used a similar procedure as described in Reference \cite{Khasanov2007}.
To analyse the data, obtained under hydrostatic pressure, the following function was added: 
\begin{equation}
\begin{aligned}
P_{pc}(t)= A_{pc}\exp\Big[-\frac{\sigma_{pc}^2t^2}{2}\Big]\cos(\gamma_{\mu}B_{int,pc}t+\varphi), 
\end{aligned}
\end{equation}
Here $A_{\rm pc}$  denote the initial asymmetry of the pressure cell.  The Gaussian relaxation rate, ${\sigma}_{\rm pc}$, reflects the depolarization due to the nuclear moments of the pressure cell. The width of the pressure cell signal increases below the critical temperature $T_{\rm c}$. As shown previously \cite{MaisuradzePC}, this is due to the influence of the diamagnetic moment of the superconducting sample on the pressure cell, leading to the temperature dependent ${\sigma}_{\rm pc}$ below the critical temperature $T_{\rm c}$. In order to consider this influence, we assume the linear coupling between ${\sigma}_{\rm pc}$ and the field shift of the internal magnetic field in the superconducting state: ${\sigma}_{\rm pc}$($T$) = ${\sigma}_{\rm pc}$($T$ \textgreater $T_{\rm c}$) + $C(T)$(${\mu}_{\rm 0}$$H_{\rm int,NS}$ - ${\mu}_{\rm 0}H_{\rm int,SC}$), where ${\sigma}_{\rm pc}$($T$ {\textgreater} $T_{\rm c}$) = 0.25 ${\mu} s^{-1}$ is the temperature independent Gaussian relaxation rate. ${\mu}_{\rm 0}$$H_{\rm int,NS}$ and ${\mu}_{\rm 0} H_{\rm int,SC}$ are the internal magnetic fields measured in the normal and in the superconducting state, respectively. The TF-$\mu$SR spectra of all samples show a strong relaxation of the signal below $T_{\rm c}$, which is evidence for the bulk type-II superconductivity in these compounds and also for CaAlSi under $p$ = 1.9 GPa.

\begin{figure}
\centering
\includegraphics[width=0.7\linewidth]{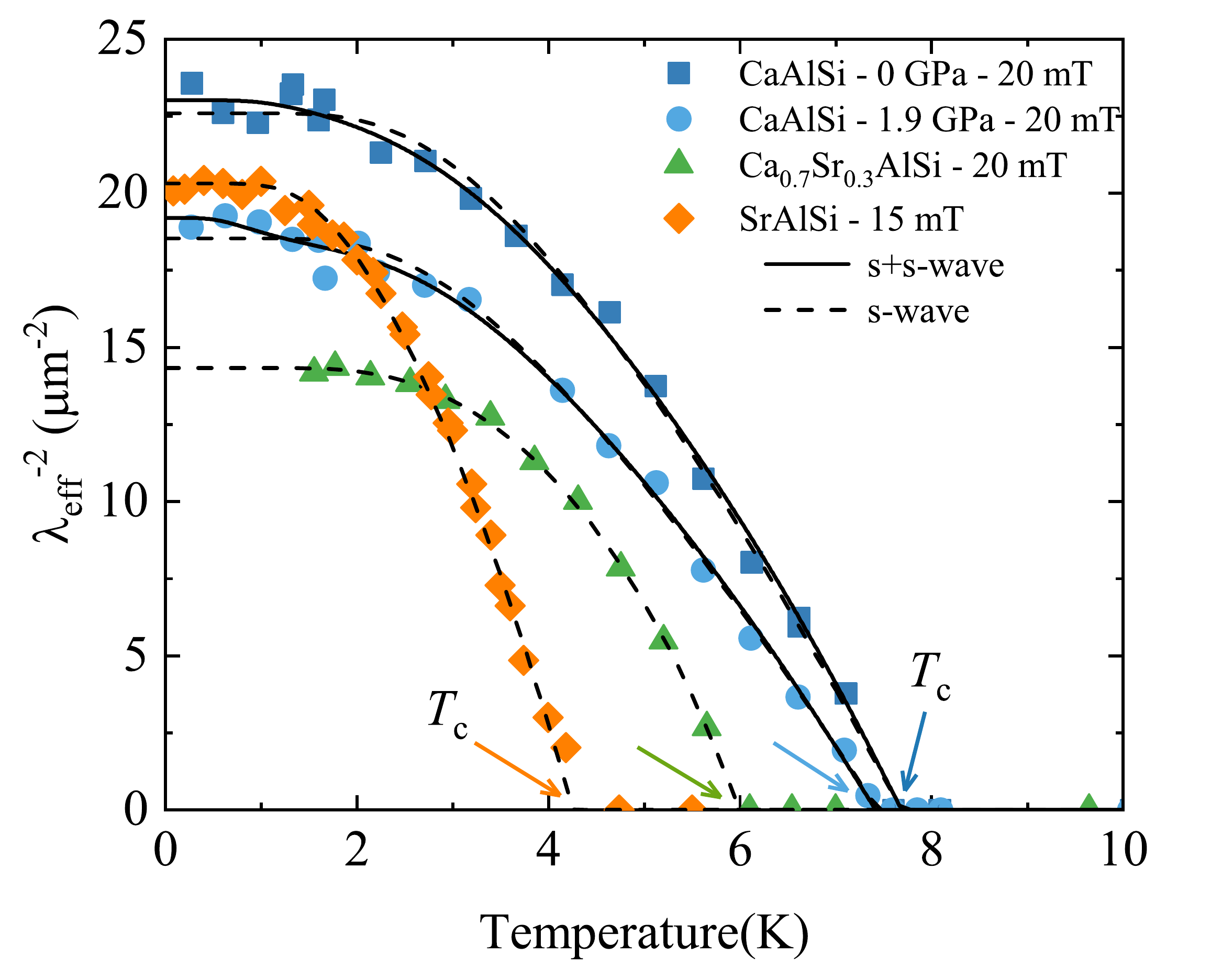}
\vspace{-0.5cm}
\caption{Temperature evolution of the London magnetic penetration depth ${\lambda}^{-2}$($T$) for \ce{Ca_{1-x}Sr_{x}AlSi}. (a) The temperature dependence of the inverse squared magnetic penetration depth ${\lambda}^{-2}$ fo CaAlSi (blue), Ca$_{0.7}$Sr$_{0.3}$AlSi (green), and SrAlSi (orange), measured at ambient and under pressure in a temperature range between $T =$ 300 mK and 10 K. Dashed line represents a fit with the single gap $s$-wave model, while the solid line correspond to the two-gap $s+s$-wave model.}
\label{fig:muon1}
\end{figure}

From the TF-${\mu}$SR spectra of \ce{Ca_{1-x}Sr_{x}AlSi} with $x$ = 0, 0.3, and 1, the superconducting part of the square root of the second moment of the field distribution in the vortex state \cite{Brandt1988,Brandt2003} ($\sqrt{\left<\Delta B^2\right>}$ ${\propto}$ ${\sigma}_{\rm sc}$ ${\propto}$ $\lambda_{eff}^{-2}$) was obtained by subtracting the nuclear moment contribution (${\sigma}_{\rm nm}$) measured at $T$ ${\textgreater}$ $T_{\rm c}$ according to ${\sigma}_{\rm sc}^{2}$ = ${\sigma}^{2}$ - ${\sigma}_{\rm nm}^{2}$. $\sigma_{sc}$ is the Gaussian relaxation rate due to the formation of the FLL \cite{Brandt1988}. In the analysis ${\sigma}_{\rm nm}$ was assumed to be constant over the entire temperature range.

In order to investigate the symmetry of the superconducting gap, we have therefore derived the temperature-dependent London magnetic penetration depth ${\lambda}(T)$, which is related to the relaxation rate by: 
\begin{equation}
\frac{\sigma_{sc}(T)}{\gamma_{\mu}}=0.06091\frac{\Phi_{0}}{\lambda^{2}(T)}.
\end{equation}

Here, ${\gamma_{\mu}}$ is the gyromagnetic ratio of the muon, and ${\Phi}_{{\rm 0}}$ is the magnetic-flux quantum. In figure \ref{fig:muon1}, we show the temperature dependence of the ${\lambda}^{-2}$ for \ce{Ca_{1-x}Sr_{x}AlSi} with $x$ = 0, 0.3, and 1 at ambient pressure. In addition, the ${\lambda}^{-2}$($T$) is shown for CaAlSi under the maximum applied pressure of $p =$ 1.9 GPa, using a high-pressure piston-cylinder set-up\cite{Khasanov2016}. ${\lambda}$($T$) was calculated within the local (London) approximation (${\lambda}$ ${\gg}$ ${\xi}$) by the following expression \cite{Suter2012,Tinkham}:
\begin{equation}
\begin{aligned}
\frac{\lambda^{-2}(T,\Delta_{0,i})}{\lambda^{-2}(0,\Delta_{0,i})}= & 1+\frac{1}{\pi} \int_{0}^{2\pi} \int_{\Delta(_{T,\varphi})}^{\infty} & (\frac{\partial f}{\partial E})\frac{EdEd\varphi}{\sqrt{E^2-\Delta_i(T,\varphi)^2}},
\end{aligned}
\end{equation}
where $f=[1+\exp(E/k_{\rm B}T)]^{-1}$ is the Fermi function, ${\varphi}$ is the angle along the Fermi surface, and ${\Delta}_{i}(T,{\varphi})={\Delta}_{0,i}{\Gamma}(T/T_{\rm c})g({\varphi}$)
(${\Delta}_{0,i}$ is the maximum gap value at $T=0$). The temperature dependence of the gap is approximated by the expression ${\Gamma}(T/T_{\rm c})=\tanh{\{}1.82[1.018(T_{\rm c}/T-1)]^{0.51}{\}}$ \cite{Carrington2003}, while $g({\varphi}$) describes the angular dependence of the gap and it is replaced by 1 for both an $s$-wave and an $s$+$s$-wave gap. The results of the gap analysis are shown in figure \ref{fig:muon1}.\\

The form of the temperature dependence of ${\lambda}^{-2}$ shows saturation at low temperatures for all 3 samples and also for CaAlSi at 1.9 GPa. Thus, the flat $T$-dependence of ${\lambda}^{-2}$ observed in \ce{Ca_{1-x}Sr_{x}AlSi} for low temperatures is consistent with a nodeless superconductor, in which ${\lambda}^{-2}\left(T\right)$ reaches its zero-temperature value exponentially. 

We find that the London magnetic penetration depth ${\lambda}^{-2}$($T$) for CaAlSi, at ambient pressure, as well as at a pressure of $p =$ 1.9 GPa, are best described by a two-gap $s+s$-wave symmetric superconducting gap (solid black line). Both the ambient pressure, as well as the $p =$ 1.9 GPa measurements are insufficiently described by a single-gap $s$-wave symmetric gap (dotted line).

The London magnetic penetration depth ${\lambda}^{-2}$($T$) for the samples Ca$_{0.7}$Sr$_{0.3}$AlSi and SrAlSi are, on the other hand, best described by the single-gap $s$-wave model. These results strongly indicate that strontium-doping not only stabilizes the simpler \ce{AlB2}-structure, and reduces the critical temperature, but also shifts the system from two-gap to single-gap superconductivity. Our results show that the lower critical temperatures $T_{\rm c}$ are anti correlated with two-gap $s+s$-wave superconductivity and that there is a critical value of $T_{\rm c}$ which separates two types of superconductivity in the solid solution of Ca$_{1-x}$Sr$_{x}$AlSi. Noteworthy is also the effect of the pressure for the CaAlSi sample. Applying 1.9 GPa has a big effect on the superfluid density, $\lambda_{eff}^{-2}$ which changes from 23.0 $\mu m^{-2}$ to 19.1 $\mu m^{-2}$ under pressure, whereas it has a small effect on the transition temperature (change from 7.7 K to 7.4 K). 


\begin{figure*}
\centering
\includegraphics[width=1.0\linewidth]{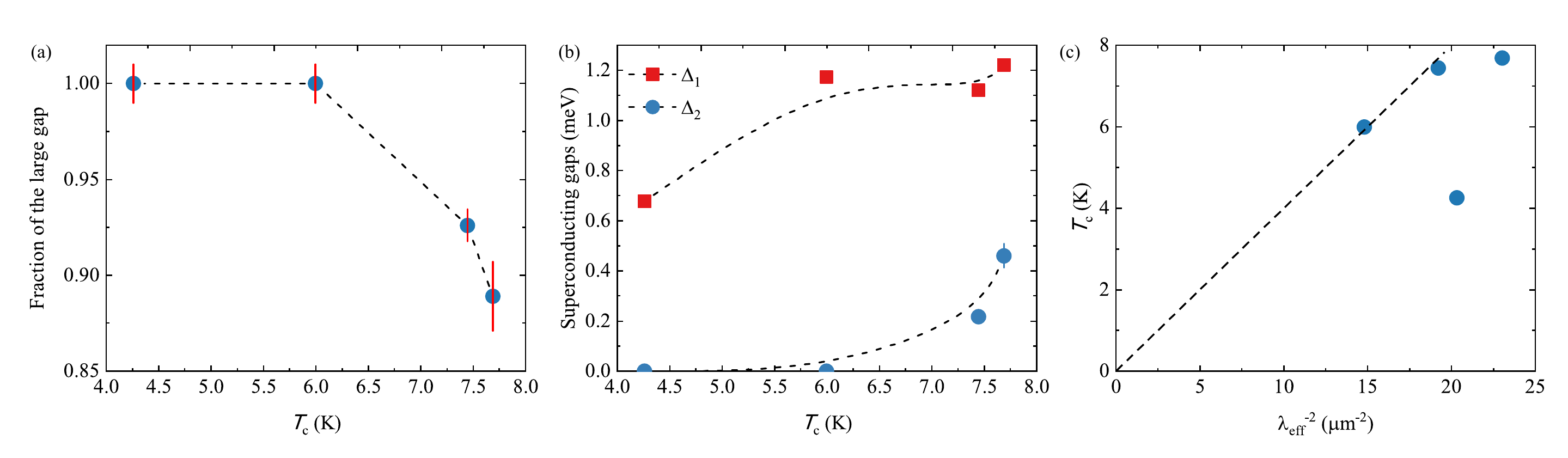}
\vspace{-0.5cm}
\caption{Superconducting gap evolution as a function of the critical temperature. Details of the analysis of ${\lambda}^{-2}$ and $\Delta_0$ for $x$ = 0, 0.3, and 1 with (a) the fraction of the large gap vs $T_{\rm c}$, (b) the size of the superconducting gaps vs $T_{\rm c}$, and (c) the critical temperature $T_{\rm c}$ against the ${\lambda}^{-2}(0)$ obtained from the ${\mu}$SR experiments. In figures (a)\&(b) the dotted line is a guide for the eye, while in figure (c) it represents the expected behavior according to the BCS theory.}
\label{fig:muon2}
\end{figure*}

In the figures \ref{fig:muon2}(a)\&(b), we show the fraction of the large gap and the values of large superconducting gap ${\Delta}_{1}$ as well as small gap ${\Delta}_{2}$, respectively, as a function of $T_{\rm c}$. These results suggest the critical value for $T_{\rm c,cr}$ ${\simeq}$ 6 K, above which the fraction of the second superconducting gap with the gap value of ${\Delta}_{2}$ = 0.2 meV starts to evolve. Hence, the substitution of calcium with strontium and the application of pressure can be interpreted as tuning parameters that shift the critical temperature $T_{\rm c}$ below the critical value, and move the system towards a single gap superconductivity. 

Besides the transition between $s$- to $s+s$-wave superconductivity, another important observation from this data is that there is an unconventional effect on the superfluid density $\rho_{\rm S}$ = $n_{s}/m^{*}$ (where $n_{s}$ is the superconducting carrier density and $m^{*}$ is the effective mass) by strontium doping and the application of pressure. This becomes apparent when we compare the extrapolated values of the London magnetic penetration depth at absolute zero $\lambda_{eff}^{-2}$(0) with each other. $\lambda_{eff}^{-2}$(0) decreases from a value of 23.0 $\mu m^{-2}$ for CaAlSi with $T_{\rm c}$ = 7.7 K to 14.3 $\mu m^{-2}$ for Ca$_{0.7}$Sr$_{0.3}$AlSi with $T_{\rm c}$ = 6.0 K. The correlation between  $\lambda_{eff}^{-2}$ and $T_{\rm c}$ is consistent with the behavior observed in other unconventional two-gap superconductors  \cite{Muechler2019,VonRohr2019,Guguchia2017,Prozorov2006a,Khasanov2008}. However, for SrAlSi, in which $T_{\rm c}$ is further reduced to 4.2 K, then $\lambda_{eff}^{-2}$(0) significantly increases to 20.3 $\mu m^{-2}$. This means that for SrAlSi, a much higher superfluid density is observed than it is expected from the value of the critical temperature, which is normally the case for conventional superconductors (compare, e.g. references \cite{VonRohr2013,Guguchia2019WS2}). 
This non-monotonous behavior of $\lambda_{eff}^{-2}$(0) and its strong enhancement for SrAlSi supports the idea that the nature of superconductivity changes depending whether the compound displays a two-gap or single-gap superconducting state. 

\section{Summary and Conclusion}
\label{sec:conclusion}

We successfully synthesized the Ca\textsubscript{1-x}Sr\textsubscript{x}AlSi solid solution with $x$ = 0, 0.05, 0.1, 0.15, 0.2, 0.25, 0.3, 0.4 0.5, 0.6, 0.7, 0.8, 0.9, and 1. We have analyzed the structure and phase purity by X-ray diffraction. We show by employing single-crystal X-ray diffraction that the structure of CaAlSi consists of a 6-folded unit cell along the $c$-axis with none of the honeycomb Al/Si layers being planar, but all of them slightly buckled. Hence, CaAlSi crystallizes in its own unique structure in the space group \textit{P}6\textsubscript{3}\textit{/mmc}. We have found all other members of the solid solution to crystallize in the \ce{AlB2}-type structure with random site occupancies in the Al/Si and Ca/Sr layers, respectively. We have, furthermore, shown that all members of the solid solution are bulk superconductors with monotonically varying critical temperatures $T_{\rm c}$ by means of magnetization measurements. The highest superconducting transition is observed for the parent compound of the solid solution $x$ = 0 (CaAlSi), with a critical temperature of $T_{\rm c} \approx$ 7.8 K. We find the critical temperature to monotonically decrease with increasing strontium content, reaching a value of $T_{\rm c} \approx$ 4.9 K for $x$ = 1 (SrAlSi). 
We found that while the cell parameters $a$ and $c$ change across the solid solution nearly linearly, following Vegard's law, the critical temperature $T_{\rm c}$ has a more complex change as a function of strontium content. The critical temperature decreases more drastically close to the structurally distorted end member CaAlSi. This indicates that the higher $T_{\rm c}$ of CaAlSi and compositions close to it, might be affected by an enhanced electron-phonon coupling due to a phonon-softening close to the structural instability. 
This is further supported by an analysis of the microscopic superconducting properties of the solid solution namely the measurements of the London magnetic penetration depth by means of $\mu$SR measurements. We have shown that SrAlSi possesses a magnetic penetration depth $\lambda_{eff}^{-2}$(0) of 20.3 $\mu m^{-2}$, while for Ca\textsubscript{0.7}Sr\textsubscript{0.3}AlSi it is equal to 14.3 $\mu m^{-2}$. Both compounds are one gap superconductors with gaps of $\Delta_0$ = 0.68 meV and 1.17 meV, respectively. CaAlSi has a magnetic penetration depth of 23.0 $\mu m^{-2}$ and is a two gap superconductor with gaps of $\Delta_1$ = 1.22 meV and $\Delta_2$ =  0.46 meV. The London penetration depth changes to 19.1 $\mu m^{-2}$ and gaps to $\Delta_1$ = 1.12 meV and $\Delta_2$ = 0.22 meV, respectively, under 1.9 GPa pressure, which indicates that pressure pushes system towards the one gap model. We find that the subtle change in the crystal structure, i.e. the buckling of the layers in CaAlSi, does not only enhance the critical temperature in the \ce{Ca_{1-x}Sr_xAlSi} solid solution substantially, but that it also initiates a single-gap to two-gap superconducting transition, leading to unconventional superconducting properties in the end member of the solid solution. The London magnetic penetration depth ${\lambda}^{-2}$($T$) for CaAlSi, at $p$ = 0 GPa and 1.9 GPa, is well described by a two-gap $s+s$-wave model, while ${\lambda}^{-2}$($T$) for the samples \ce{Ca_{0.7}Sr_{0.3}AlSi} and SrAlSi are compatible with the single-gap $s$-wave model. We have shown that the system \ce{Ca_{1-x}Sr_xAlSi} is a rich superconducting system, where a structural transition and a two-gap to single-gap superconducting transition can be controlled by a isoelectronic chemical substitution, or by pressure, making this a most promising model system for the investigation of the entanglement of structural and electronic interactions in superconductors on honeycomb lattices. Our results may contribute to a better understanding of structure-composition-property relations in layered superconducting materials in general.

\section{Acknowledgements}

We thank Markus Bendele for experimental help during the beginning of this project, and Andreas Schilling for helpful discussions. This work was supported by the Swiss National Science Foundation under Grant No. PZ00P2\_174015.

\bibliography{apssamp}

\end{document}